\documentclass[10pt]{article}

\usepackage{fullpage}
\usepackage{setspace}
\usepackage{parskip}
\usepackage{titlesec}
\usepackage{placeins}
\usepackage{xcolor}
\usepackage{lineno}
\usepackage[toc,page]{appendix}
\usepackage{physics}
\usepackage[numbers,sort&compress]{natbib}

\PassOptionsToPackage{hyphens}{url}
\usepackage[colorlinks = true,
            linkcolor = blue,
            urlcolor  = blue,
            citecolor = blue,
            anchorcolor = blue]{hyperref}
\usepackage{etoolbox}
\makeatletter
\patchcmd\@combinedblfloats{\box\@outputbox}{\unvbox\@outputbox}{}{%
}%
\makeatother

\renewenvironment{abstract}
  {{\bfseries\noindent{\large\abstractname}\par\nobreak}}

\titlespacing{\section}{0pt}{*3}{*1}
\titlespacing{\subsection}{0pt}{*2}{*0.5}
\titlespacing{\subsubsection}{0pt}{*1.5}{0pt}

\usepackage{authblk}

\usepackage{graphicx}
\usepackage[space]{grffile}
\usepackage{latexsym}
\usepackage{textcomp}
\usepackage{longtable}
\usepackage{tabulary}
\usepackage{booktabs,array,multirow}
\usepackage{amsfonts,amsmath,amssymb}
\providecommand\citet{\cite}
\providecommand\citep{\cite}

\newif\iflatexml\latexmlfalse
\DeclareGraphicsExtensions{.pdf,.PDF,.png,.PNG,.jpg,.JPG,.jpeg,.JPEG}

\usepackage[utf8]{inputenc}
\usepackage[english]{babel}

\begin{document}

\title{A Laboratory Frame Density Matrix for Ultrafast Quantum Molecular Dynamics }

\author[1]{Margaret Gregory
\thanks{Current Address: Department of Earth, Atmospheric, and Planetary Sciences, Massachusetts Institute of Technology, 77 Massachusetts Avenue, Cambridge, MA 02139, USA.}}
\author[2]{Simon Neville}
\author[2,3]{Michael Schuurman}
\author[1]{Varun Makhija
\thanks{vmakhija@umw.edu}}
\affil[1]{Department of Chemistry and Physics, University of Mary Washington, 1301 College Avenue, Fredericksburg VA, 22401.}%
\affil[2]{National Research Council Canada, 100 Sussex Drive, Ottawa, ON, K1A 0R6, Canada}%
\affil[3]{Department of Chemistry and Biomolecular Sciences, University of Ottawa, 150 Louis Pasteur, Ottawa, ON, K1N 6N5, Canada}

\vspace{-1em}

  \date{\today}

\begingroup
\let\center\flushleft
\let\endcenter\endflushleft
\maketitle
\endgroup

\begin{abstract}
    In most cases the ultrafast dynamics of resonantly excited molecules are considered, and almost always computed in the molecular frame, while experiments are carried out in the laboratory frame. Here we provide a formalism in terms of a lab frame density matrix which connects quantum dynamics in the molecular frame to those in the laboratory frame, providing a transparent link between computation and measurement. The formalism reveals that in any such experiment, the molecular frame dynamics vary for molecules in different orientations and that certain coherences which are potentially experimentally accessible are rejected by the orientation-averaged reduced vibronic density matrix. Instead, Molecular Angular Distribution Moments (MADMs) are introduced as a more accurate representation of experimentally accessible information. Furthermore, the formalism provides a clear definition of a molecular frame quantum tomography, and specifies the requirements to perform such a measurement enabling the experimental imaging of molecular frame vibronic dynamics. Successful completion of such a measurement fully characterizes the molecular frame quantum dynamics for a molecule at any orientation in the laboratory frame.  
\end{abstract}

\section{Introduction}

Discerning the dynamics of resonantly excited isolated molecules is a challenging many body quantum problem, which has spurred the development of numerous computational and experimental methods in chemistry and physics~\cite{kouppel1984, felker1985, williamson1997, beck2000, ben2000, jonas2003, domcke2004, stolow2004, suzuki2006, richter2011, Reid2012, timmers2014, beck2015, minitti2015, glownia2016, neville2016, smith2018, kirrander2016,smith2018, bellshaw2019,wolf2019,li2020,zinchenko2021,karamatskos2021,razmus2022}. The combination of these powerful experimental and theoretical techniques can provide reduced dimensionality models that explain spectroscopic observations. Typically, studies of femtosecond and attosecond dynamics focus on vibrational and/or electronic dynamics and ignore the rotational degrees of freedom since rotation occurs on much longer timescales~\cite{nisoli2017,rouxel2018,neville2018,despre2018,hermann2020,folorunso2021}. Vibrational and electronic motion are also of more immediate interest, since their coupling drives the transfer of energy in molecular systems~\cite{kouppel1984, domcke2004, stolow2004}. In particular, the development of attosecond laser pulses provides the opportunity to directly measure electronic coherences prior to dephasing~\cite{sansone2006,calegari2014,suominen2014, kowalewski2015,nisoli2017,arnold2018, simmermacher2019,duris2020,blavier2021,carrascosa2021,giri2021,tremblay2021, ruberti2021, neville2022,mauger2022,li2022}, generating much recent interest in directly imaging charge transfer dynamics~\cite{rouxel2018,despre2018,hermann2020,folorunso2021}. However, electronic and vibrational coordinates are defined in the molecular frame, and experiments always occur in the laboratory frame. We may greatly enhance the probability of a particular molecular orientation for some molecules using laser-alignment techniques~\cite{stapelfeldt2003,koch2019}, but quantum mechanics precludes the possibility of producing a fixed orientation~\cite{mayer1996,grohmann2021}. In other words, considering the vibronic (vibrational-electronic) dynamics in a fixed molecule is an approximation. We may then ask what, if any, are the measurable consequences of this approximation? In particular, does this approximation impact the efficacy of theoretical proposals for measuring coherent molecular dynamics?  Quantum mechanically, ignoring rotation amounts to neglecting the manifold of very closely spaced total angular momentum eigenstates that lie on each vibronic  state. The bandwidth of ultrashort laser pulses invariably exceeds this energy gap, exciting a large superposition of angular momentum states. Here, we develop a density matrix-based framework for vibronic dynamics that accounts for this superposition, thus including the rotational degrees of freedom. This framework can be considered a straight forward extension of that developed by Fano to study the alignment and orientation of electrons in excited atoms, which has also been applied for this purpose to diatomic molecules~\cite{fano1972,blum2012}. 

Even though rotational dynamics do not occur on femto and attosecond timescales, we find that molecules at different orientations in the lab frame will exhibit different vibronic dynamics. Additionally, certain coherences that may appear in a fixed molecule calculation can be explicitly rejected by the reduced vibronic density matrix (with rotations traced out or, equivalently, orientation-averaged). This may be particularly relevant for experiments that target specific density matrix elements, which is sometimes possible by two-dimensional spectroscopy~\cite{jonas2003}. Further, as we discuss in detail, these missing coherences can potentially be probed experimentally making the orientation averaged vibronic density matrix an inaccurate representation of such experiments. Finally, since molecules at different orientations undergo different dynamics, we show that it is potentially possible to extract density matrix elements from an experiment in which vibronic dynamics are initiated from initially \emph{un-oriented} molecules. In particular cases it may be possible to extract all relevant density matrix elements experimentally, thus completely determining the molecular frame vibronic dynamics for a molecule at any orientation in the lab frame. This therefore constitutes a molecular frame quantum tomography.     

We derive and describe the density matrix framework in section~\ref{sec:framework} below. A core component of the framework are the Angular Momentum Coherence Operators (AMCOs) introduced in section~\ref{sec:AMCO}. These are an extension of the state multipoles introduced by Fano~\cite{fano1972,blum2012}, however we do not endeavour to explore the tensor nature of these operators in detail. The AMCOs then facilitate the decomposition of a lab frame density matrix derived in section~\ref{sec:density} into time-depdendent Molecular Angular Distribution Moments (MADMs), which encode the orientation dependent vibronic dynamics of the molecule in the lab frame. The MADMs are also the quantities that provide a direct connection between the density matrix and measurement. In section~\ref{sec:qtomography} this connection is explored in detail for time resolved photoelectron angular distributions from a resonantly excited molecule. Making a sufficient number of measurements such that all relevant MADMs can be determined leads to a molecular frame quantum tomography. Finally, we conclude by providing a summary and discuss some potential applications of our work.

\section{Theoretical Framework}
\label{sec:framework}
\subsection{Preliminaries}

We first consider vibrational and electronic dynamics in the molecular frame. These are typically described in a Born-Oppenheimer (BO) basis of electronic and vibrational states, which we label $\ket{\alpha}$ and $\ket{v_\alpha}$, respectively. A broadband light pulse can excite a superposition of these states. Typically, it is assumed that the excitation occurs out of a single state, allowing the resulting wavepacket to be written as a pure state,
\begin{equation}
    \ket{\Psi_{mol}(t)}=\sum_{\alpha v_\alpha} C_{\alpha v_\alpha}(t) \ket{\alpha} \ket{v_\alpha}.
    \label{eq:Psimol}
\end{equation}
Ignoring rotation implicitly restricts us to the molecular frame, hence the state label $\ket{\Psi_{mol}(t)}$. The resulting molecular wavefunction can be found by projecting onto the electronic positions and vibrational modes basis, which we label $\ket{\textbf{r}}$ and $\ket{\textbf{Q}}$, respectively. Here $\textbf{r}=\{r_i|i=1,2,\dots,\}$ represents the set of electronic position vectors in the molecular frame, and $\textbf{Q}=\{q_1,q_2,q_3,\dots\}$ the set of vibrational coordinates. Thus, the molecular probability distribution can be expressed as follows,
\begin{equation}
    |\Psi_{mol}(t, \textbf{r},\textbf{Q})|^2=|\bra{\textbf{r},\textbf{Q}}\ket{\Psi_{mol}(t)}|^2=\sum_{\alpha \alpha' v_\alpha v'_{\alpha'}} \Phi_\alpha(\textbf{r}) \Phi^*_{\alpha'}(\textbf{r})\psi_{v_\alpha}(\textbf{Q}) \psi^*_{v'_{\alpha'}}(\textbf{Q}) C_{\alpha, v_\alpha}(t)C^*_{\alpha', v'_{\alpha'}}(t)
    \label{eq:Psisquared}
\end{equation}
where $\Phi_\alpha(\textbf{r})=\bra{\textbf{r}}\ket{\alpha}$ is the electronic wavefunction for electronic state $\ket{\alpha}$, with $\abs{\Phi_\alpha(\textbf{r})}^2$ the probability density of finding an electron located at $\textbf{r}$ in the state $\ket{\alpha}$. Similarly, $\psi_{v_\alpha}(\textbf{Q})=\bra{\textbf{Q}}\ket{v_\alpha}$ is the vibrational wavefunction for vibrational state $\ket{v_\alpha}$, with $\abs{\psi_{v_\alpha}(\textbf{Q})}^2$ the probability density of locating the nuclei at $Q$ in the state $\ket{v_\alpha}$.  
Rewriting Eq.~\ref{eq:Psisquared} in terms of the density matrix and separating it into diagonal and off-diagonal terms:
\begin{equation}
\begin{split}
     |\Psi_{mol}(t, \textbf{r}, \textbf{Q})|^2=\sum_{\alpha,v_{\alpha}} \rho(\alpha, v_\alpha,t) P_{\alpha, v_\alpha}(\textbf{r}, \textbf{Q}) +\\ \sum_{\alpha \neq \alpha'} \sum_{v_{\alpha} \neq v'_{\alpha'}} \rho(\alpha, v_\alpha, \alpha',v'_{\alpha'},t)\Phi_\alpha(\textbf{r}) \Phi^*_{\alpha'}(\textbf{r})\psi_{v_\alpha}(\textbf{Q}) \psi^*_{v'_{\alpha'}}(\textbf{Q}) 
    \label{eq:psimolsquared}
\end{split}
\end{equation}
In the equation above, $\rho(\alpha, v_\alpha,t)=\abs{C_{\alpha, v_\alpha}(t)}^2$ and $\rho(\alpha, v_\alpha, \alpha',v'_{\alpha'},t)=C_{\alpha, v_\alpha}(t)C^*_{\alpha', v'_{\alpha'}}(t)$ are the diagonal and off-diagonal elements of the density matrix. Additionally, $P_{\alpha v_\alpha}(\textbf{r},\textbf{Q})=\abs{\Phi_\alpha(\textbf{r})}^2\abs{\psi_{v_\alpha}(\textbf{Q})}^2$ is the probability density of locating the electrons at $\textbf{r}$ and the nuclei at $Q$ in the Born-Oppenheimer State $\ket{\alpha v_\alpha}$. Thus, the values of $\rho(\alpha, v_\alpha,t)$ give the contribution of a particular BO State to the total molecular probability density in the molecular frame, and are the so-called populations. The term $\Phi_\alpha(\textbf{r}) \Phi^*_{\alpha'}(\textbf{r})$ gives the spatial overlap between $\Phi_\alpha(\textbf{r})$, and $\Phi_{\alpha'}(\textbf{r})$ and $\psi_{v_\alpha}(\textbf{Q}) \psi^*_{v'_{\alpha'}}(\textbf{Q})$ describes the spatial overlap between $\psi_{v_\alpha}(\textbf{Q})$ and $\psi_{v'_{\alpha'}}(\textbf{Q})$. Hence, the coherences $\rho(\alpha, v_\alpha, \alpha',v'_{\alpha'},t)$ indicate to what degree the interference between two states contributes to the total probability density. This molecular frame density matrix, or some variant thereof, is typically used to compute observables and interpret experimental data.

In subsequent sections we develop the tools needed to derive an equivalent, laboratory frame density matrix which fully characterizes the dynamics in any given experiment. This is done by including rotation and considering the total angular momentum of the molecule. To do so, the angular momenta associated with all states excited by the light pulse need to be correctly coupled. For this purpose, we introduce Angular Momentum Coherence Operators in the next subsection.

\subsection{Angular Momentum Coherence Operators}
\label{sec:AMCO}
We begin our analysis by considering the irreducible representations of the rotation group SO(3), the Wigner matrix elements $D^j_{mk}$~\cite{zare1988,sakurai}. Here $j$ specifies the representation, with $m$ and $k$ running from $-j$ to $j$ facilitating a rotation between basis functions within a representation~\cite{zare1988,sakurai},
\begin{equation}
\left|jm\right\rangle=\sum_{k}D^{j*}_{mk}(\mathbf{\Omega})\left|jk\right\rangle,
\end{equation}	  
$\mathbf{\Omega}$ specifying the rotation angles, here taken to be the Euler angles $\mathbf{\Omega}=\{\phi,\theta,\chi\}$. The properties of the $D^{j}_{mk}$ make them suitable basis functions for molecular rotation. Of particular note is that the irreducible representations $j_1$ and $j_2$ are easily combined to generate irreducible representations $j_3$~\cite{zare1988,sakurai},
\begin{equation}
 D^{j_1}_{m_1k_1}(\Omega)D^{j_2}_{m_2k_2}(\Omega)=\sum_{j_3}(2j_3 +1)\left(\begin{array}{c c c} j_1 & j_2 & j_3 \\ m_1 & m_2 & m_3 \end{array}\right) \left(\begin{array}{c c c} j_1 & j_2 & j_3 \\ k_1 & k_2 & k_3 \end{array}\right) D^{j^{*}_3}_{m_3k_3}(\Omega)
\label{eq:DDcombination}
\end{equation}
where the large parenthesis are the Wigner three-j symbols. This enables straightforward coupling of angular momenta such that $j_3 = |j_1-j_2|,|j_1-j_2+1|,\dots,|j_1+j_2-1|,|j_1+j_2|$, $m_3=m_1+m_2$ and $k_3=k_1+k_2$. Furthermore the eigenfunctions, $\left|JKM\right\rangle$, of the rigid symmetric top Hamiltonian are also given by the Wigner matrix elements~\cite{zare1988}, 
\begin{equation}
\left\langle\mathbf{\Omega}|JKM\right\rangle=\sqrt{\frac{2J+1}{8\pi^2}}D^{J*}_{MK}(\mathbf{\Omega}).
\label{eq:JKMbasis}
\end{equation} 
Here $J$ is the total angular momentum, $K$ its projection on a chosen molecular axis and $M$ on the space fixed axis. The asymmetric top Hamiltonian mixes states of different $K$, but $J$ is conserved. In fact no intramolecular interaction can mix states of different $J$ since the rotation group SO(3) is a symmetry group of the full molecular Hamiltonian - the total angular momentum $J$ conserved. With these considerations it is appropriate to specify orthogonal molecular basis functions $\left|n\right\rangle \left |J_nK_nM_n\right\rangle$, with $n = {\alpha,v_\alpha}$ labeling molecular frame vibronic BO basis states used above. While intramolecular interactions conserve angular momentum, a polarized laser pulse necessarily excites a coherent superposition of $J_n$ states. These can easily be coupled as prescribed by Eq.~\ref{eq:DDcombination} to give the total angular momentum of a wavepacket of states. This motivates definition of the following tensor operators, indexed by the total angular momentum and its projections,
\begin{equation}
\begin{split}
&A^{K}_{QS}=\frac{2K+1}{8\pi^2}\sum_{J_n K_n M_n}\sum_{J'_{n^\prime} K'_{n^\prime} M'_{n^\prime}}\sum_{n n'}\sqrt{(2J_n+1)(2J'_{n'}+1)}(-1)^{M_{n}-K_{n}}\\
&\times \left( \begin{array}{ccc} J_n & J'_{n\prime} & K \\ -M_n & M'_{n'} & Q \end{array} \right) 
\left( \begin{array}{ccc} J_n & J'_{n'} & K \\ -K_n & K'_{n'} & S \end{array} \right) \left|n'\right\rangle \left|J'_{n'} K'_{n'} M'_{n'}\right\rangle \left\langle J_n K_n M_n\right| \left\langle n\right|.
\label{eq:AKQS}
\end{split}
\end{equation}   
Here $K = |J_{n}-J'_{n'}|,|J_{n}-J_{n'}+1|,\dots,|J_{n}+J_{n'}-1|,|J_{n}+J_{n'}|$ is the total wavepacket angular momentum which results from coupling the angular momenta in the excited wavepacket. We adhere to the unfortunate convention where $K$ is the total wavepacket angular momentum while $K_n$ is the projection of $J_n$ onto the molecular axis. $Q$ and $S$ are the projections of $K$ onto the space fixed and molecular axes respectively~\cite{underwood2008}. Note that these are constructed from differences of the projection angular momenta - $Q=M'_{n'}-M_{n}$ and $S=K'_{n'}-K_{n}$ - rather than the sum as in Eq.~\ref{eq:DDcombination}. This is done so that the set of operators $A^K_{QS}$ project out quantum beats between states of different angular momentum, which in turn correspond to a time varying anisotropy as shown below. We refer to the $A^K_{QS}$ as Angular Momentum Coherence Operators (AMCOs). They are defined in a manner similar to the state multipole operators $T^{K}_{Q}$ introduced by Fano and used to analyze anisotropy in coherent distributions of $M_n$ states. As we will show below the AMCOs, as defined above, are an important part of calculating the matrix elements of the density operator and understanding the origins and behavior of coherences in the laboratory frame. 
 
 \subsection{The Laboratory Frame Density Matrix}
 \label{sec:density}
 Matrix elements of the density operator, like the molecular wavefunction, can be used to determine all observable properties of a molecule~\cite{blum2012}. In the $\ket{n}\ket{J_nK_nM_n}$ basis, the density matrix has elements 
 \begin{equation}
     \bra{J'_{n'}K'_{n'}M'_{n'}n'}\hat{\rho}(t)\ket{nJ_nK_nM_n}=\rho_{\omega_n \omega'_{n'}}(n,n',t)
     \label{eq:rho}
 \end{equation}
 with $\omega_n=\{J_n K_n M_n\}$ and $n=\alpha, v_\alpha$. This constitutes a density matrix in the laboratory frame, in contrast to the molecular frame quantity in Eq.~\ref{eq:psimolsquared}. We may write the density operator in terms of the density matrix elements $\rho_{\omega_n \omega'_{n'}}(n,n',t)$, as follows,
\begin{equation}
\hat{\rho}(t)=\sum_{\alpha \alpha' v_\alpha v'_{\alpha'}}\sum_{J_n K_n M_n} \sum_{J'_{n'} K'_{n'} M'_{n'}} \rho_{\omega_n \omega'_{n'}}(n,n',t)\ket{\alpha} \ket{v_\alpha}\ket{J_n K_n M_n}\bra{J'_{n'} K'_{n'} M'_{n'}}\bra{\alpha'}\bra{v'_{\alpha'}}
\label{eq:rho-op}
\end{equation}
This equation is consistent with Eq. ~\ref{eq:rho} since the BO states and $\bra{J'_{n'}K'_{n'}M'_{n'}}\ket{J_nK_nM_n}=\delta_{J'_{n'}J_{n}}\delta_{K'_{n'}K_{n}}\delta_{M'_{n'}M_{n}}$ are both orthogonal. For a molecule with orientation given by the Euler Angles $\phi$, $\theta$, and $\chi$, in a nuclear configuration $\textbf{Q}$ the probability of finding the electrons at $\textbf{r}$ is given by the diagonal elements of $\hat{\rho}(t)$ in the  $\ket{\textbf{Q}}\ket{\textbf{r}}\ket{\phi, \theta, \chi}$ basis -  $\bra{\textbf{Q},\textbf{r};\phi, \theta, \chi}\hat{\rho}(t)\ket{\textbf{Q},\textbf{r};\phi, \theta, \chi}$. 
\begin{equation}
\begin{split}
     P(\textbf{Q},\textbf{r},\mathbf{\Omega},t)=\sum_{\alpha \alpha' v_\alpha v'_{\alpha'}}\sum_{\omega_n \omega'_{n'}}\rho_{\omega_n \omega'_{n'}}(n,n',t) \Phi_\alpha(\textbf{r}) \Phi^*_{\alpha'}(\textbf{r})\psi_{v_\alpha}(\textbf{Q}) \psi^*_{v'_{\alpha'}}(\textbf{Q})
     \\
     \times\frac{ \sqrt{(2J_n+1)(2J'_{n'}+1)}}{8\pi^2} (-1)^{M_n-K_n} D^{J_n}_{-M_n -K_n}(\mathbf{\Omega})D^{J'_{n'}}_{M'_{n'} K'_{n'}}(\mathbf{\Omega})
     \end{split}
\end{equation}
Here $P(\textbf{Q},\textbf{r},\mathbf{\Omega},t)$ is the probability of finding a molecule oriented at $\mathbf{\Omega}$ with a configuration of nuclei located at $\textbf{Q}$, and electrons located at $\textbf{r}$.  
Using Eq.~\ref{eq:DDcombination} to couple angular momenta gives
\begin{equation}
\begin{split}
    P(\textbf{Q},\textbf{r},\mathbf{\Omega},t) = \sum_{\alpha \alpha' v_\alpha v'_{\alpha'}} \Phi_\alpha(\textbf{r}) \Phi^*_{\alpha'}(\textbf{r})\psi_{v_\alpha}(\textbf{Q}) \psi^*_{v'_{\alpha'}}(\textbf{Q})\\ \times\sum_{\omega_n \omega'_{n'}}\rho_{\omega_n \omega'_{n'}}(n,n',t)
    \frac{\sqrt{(2J+1)(2J'_{n'}+1)}}{8 \pi^2} (-1)^{M_n-K_n} \sum_{K Q S} (2K+1)\\ \times \begin{pmatrix} J_n & J'_{n'} & K \\ -M_n & M'_{n'} & Q\end{pmatrix} \begin{pmatrix} J_n & J'_{n'} & K \\ -K_n & K'_{n'} & S\end{pmatrix} D^{K*}_{Q S}(\mathbf{\Omega})
    \end{split}
\end{equation}
Rewriting this equation with 
\begin{equation}
\begin{split}
    \bra{\alpha' v'_{\alpha'} \omega'_{n'}} \hat{A}^K_{Q S}\ket{\alpha v_{\alpha} \omega_{n}}=\frac{\sqrt{(2J+1)(2J'_{n'}+1)}}{8\pi^2} (-1)^{M_n-K_n} \\ \times (2K+1) \begin{pmatrix} J_n & J'_{n'} & K \\ -M_n & M'_{n'} & Q\end{pmatrix} \begin{pmatrix} J_n & J'_{n'} & K\\ -K_n & K'_{n'} & S\end{pmatrix}
    \end{split}
\end{equation}
which are the matrix elements of the AMCOs, Eq.~\ref{eq:AKQS}, in the BO basis, results in:
\begin{equation}
\begin{split}
    P(\textbf{Q},\textbf{r},\mathbf{\Omega},t)=\sum_{K Q S}\sum_{\alpha \alpha' v_\alpha v'_{\alpha'}} \Phi_\alpha(\textbf{r}) \Phi^*_{\alpha'}(\textbf{r})\psi_{v_\alpha}(\textbf{Q}) \psi^*_{v'_{\alpha'}}(\textbf{Q})
    \\
    \times \sum_{\omega_n \omega'_{n'}}\rho_{\omega_n \omega'_{n'}}(n,n',t) \bra{\alpha' v'_{\alpha'} \omega'_{n'}} \hat{A}^K_{Q S}\ket{\alpha v_{\alpha} \omega_{n}} D^{K*}_{Q S}(\mathbf{\Omega})
    \end{split}
\end{equation}
Defining the time-depdendent coefficients as:
\begin{equation}
    A^K_{Q S}(\alpha, \alpha', v_\alpha ,v'_{\alpha'};t)=\sum_{\omega_n \omega'_{n'}}\rho_{\omega_n \omega'_{n'}}(n,n',t) \bra{\alpha' v'_{\alpha'} \omega'_{n'}} \hat{A}^K_{Q S}\ket{\alpha v_{\alpha} \omega_{n}},
    \label{eq:EADMs}
\end{equation}
we get
\begin{equation}
\begin{split}
     P(\textbf{Q},\textbf{r},\mathbf{\Omega},t)=\sum_{K Q S}\sum_{\alpha \alpha' v_\alpha v'_{\alpha'}} \Phi_\alpha(\textbf{r}) \Phi^*_{\alpha'}(\textbf{r})\psi_{v_\alpha}(\textbf{Q}) \psi^*_{v'_{\alpha'}}(\textbf{Q}) \\ \times A^K_{Q S}(\alpha, \alpha', v_\alpha ,v'_{\alpha'};t) D^{K*}_{Q S}(\mathbf{\Omega})
\end{split}
\label{eq:PintermEC}
\end{equation}
Finally, we separate the diagonal terms $v_\alpha=v'_{\alpha'}$ and $\alpha=\alpha'$ from the off-diagonal terms $v_\alpha\neq v'_{\alpha'}$ and $\alpha\neq\alpha'$ and consolidate the time and orientation-angle dependent coefficients to give:
\begin{equation}
\begin{split}
     P(\textbf{Q},\textbf{r},\mathbf{\Omega},t)=\sum_{\alpha, v_\alpha}\rho_{\alpha v_\alpha}(\mathbf{\Omega},t) P_{\alpha, v_\alpha}(\textbf{r},\textbf{Q}) + \\ \sum_{\alpha \neq \alpha} \sum_{v_\alpha \neq v'_{\alpha'}}\rho_{\alpha \alpha' v_\alpha v'_{\alpha'}}(\mathbf{\Omega},t) \Phi_\alpha(\textbf{r}) \Phi^*_{\alpha'}(\textbf{r})\psi_{v_\alpha}(\textbf{Q}) \psi^*_{v'_{\alpha'}}(\textbf{Q}) 
\end{split}
\label{eq:PintermB}
\end{equation}
We can compare this with the molecular frame probability distribution Eq.~\ref{eq:psimolsquared}, which we reproduce here:
\begin{equation}
\begin{split}
 |\Psi_{mol}(t, \textbf{r}, Q)|^2=\sum_{\alpha,v_{\alpha}} \rho(\alpha, v_\alpha,t) P_{\alpha, v_\alpha}(\textbf{r}, \textbf{Q}) +\\ \sum_{\alpha \neq \alpha'} \sum_{v_{\alpha} \neq v'_{\alpha'}} \rho(\alpha, v_\alpha, \alpha',v'_{\alpha'},t)\Phi_\alpha(\textbf{r}) \Phi^*_{\alpha'}(\textbf{r})\psi_{v_\alpha}(\textbf{Q}) \psi^*_{v'_{\alpha'}}(\textbf{Q}) 
\end{split}
\label{eq:Pmol}
\end{equation}
We can thus identify  $\rho_{\alpha \alpha' v_\alpha v'_{\alpha'}}(\mathbf{\Omega},t)$ as the off-diagonal density matrix elements in the laboratory frame. Since these off diagonal matrix elements are dependent on orientation, coherences in the laboratory frame are also orientation dependent. Note in both Eq.~\ref{eq:PintermB} and Eq.~\ref{eq:Pmol}, the off-diagonal elements are coefficients of molecular frame electronic and vibrational wave functions; therefore, they track coherences between molecular frame vibrational and electronic states. This term is partially responsible for the time-depdendent nature of the probability distribution and controls which coherences in the molecular frame contribute to the laboratory frame density and at what angle and time. Similarly, we can identify $\rho_{\alpha v_\alpha}(\mathbf{\Omega},t)$ as the diagonal density matrix elements. The diagonal matrix elements give the probability of finding the molecule at an orientation $\mathbf{\Omega}$, in a BO state $\ket{\alpha v_\alpha}$ as a function of time.  

 The lab frame density matrix elements in Eq.~\ref{eq:PintermB} contain the functions $A^K_{Q S}(\alpha, \alpha', v_\alpha ,v'_{\alpha'};t)$ from Eq.~\ref{eq:EADMs} in a multipole expansion,
 \begin{equation}
     \rho_{\alpha v_\alpha}(\mathbf{\Omega},t)= \sum_{KQS} A^K_{Q S}(\alpha, v_\alpha;t) D^{K*}_{Q S}(\mathbf{\Omega})
     \label{eq:ADMOrigin}
 \end{equation}
 \begin{equation}
     \rho_{\alpha \alpha' v_\alpha v'_{\alpha'}}(\mathbf{\Omega},t)=\sum_{KQS} A^K_{Q S}(\alpha, \alpha', v_\alpha ,v'_{\alpha'};t) D^{K*}_{Q S}(\mathbf{\Omega})
     \label{eq:EADMOrigin}
 \end{equation}
 The $A^K_{Q S}(\alpha, v_\alpha;t)$ and $A^K_{Q S}(\alpha, \alpha', v_\alpha ,v'_{\alpha'};t)$ are therefore multipole moments of the diagonal and off diagonal density matrix elements, respectively. Collectively, we will refer to these as the Molecular Angular Distribution Moments (MADM). The off diagonal elements $A^K_{Q S}(\alpha, \alpha', v_\alpha ,v'_{\alpha'};t)$ determine the time varying orientation-dependent shape of the vibronic coherences in the laboratory frame, which we refer to as  Electronic Angular Distribution Moments (EADMs)~\cite{makhija2020}. The diagonal matrix elements $A^K_{Q S}(\alpha, v_\alpha;t)$ determine the time varying probability distribution of molecular orientations in a particular vibronic state, and are called the Axis Distribution Moments (ADMs) ~\cite{underwood2008}. These collectively dictate the orientation-dependent shape of the all density matrix elements in the lab frame. For example, if $K=1$ is the dominant term in the sum, the density matrix element will resemble a dipole. If the matrix element in question is diagonal, molecules in that particular vibronic state will be oriented and the $A^1_{Q S}(\alpha, v_\alpha;t)$ will track this orientation over time. If it is an off-diagonal matrix element, there will be a significant difference in the vibronic dynamics depending on the orientation of the molecule. Additionally, this difference will vary with time on a time scale determined by the energy gap between the coherently excited, molecular frame vibronic states. This time variation is encoded by $A^1_{Q S}(\alpha, \alpha', v_\alpha ,v'_{\alpha'};t)$. A familiar example is the macroscopic dipole responsible for resonant absorption induced by coherent excitation out of the ground state. Similarly, if $K=2$ is dominant then the matrix elements will resemble a quadrupole, and so on. In general, as indicated by Eq.~\ref{eq:ADMOrigin} and Eq.~\ref{eq:EADMOrigin}, the density matrix elements will be some combination of these different moments. These density matrix elements in turn determine the full-dimensional laboratory frame probability distribution of the molecule, as given by Eq.~\ref{eq:PintermB}. 
 
 The general conclusion we can draw from Eq.~\ref{eq:ADMOrigin} and Eq.~\ref{eq:EADMOrigin} is that the molecular frame dynamics will differ for molecules orientated at different angles in the lab frame, and the MADMs determine these dynamics. The lab frame density matrix elements thus provide a direct connection between the molecular frame and lab frame, and fully determine the dynamics in the molecular frame at any orientation. Eq.~\ref{eq:ADMOrigin} and Eq.~\ref{eq:EADMOrigin} indicate that a number of MADMs need to be calculated in order to determine the lab frame density matrix. In the following subsection we derive selection rules for the MADMs, restricting the sum over $K$ in these equations. We also provide relationships between the MADMs that further reduce the number that need to be computed.  
 
 \subsubsection{MADM selection Rules and relations}
 
 Angular momentum selection rules can render most MADMs zero. It is instructive to begin with the MADMs that result by one photon excitation from an isotropic, rotation-less ground state with $J = 0$, $\ket{000}$. These selection rules will also apply to one photon excitation from a mixed state representing an isotropic distribution of molecules with zero average angular momentum and zero average projection of angular momentum on any axis - the typical initial condition for most ultrafast dynamics experiments. One photon selection rules restrict excited-state angular momenta to $J_n = 1$,$K_n = 0, \pm1$ and $M_n = 0, \pm1$. In the specific case of excitation by linearly polarized light $M_n = 0$. We can now use these selection rules to determine which MADMs are non-zero upon excitation by examining the AMCOs defined by Eq.~\ref{eq:AKQS}. Matrix elements of the AMCOs then determine the MADMs $A^{K}_{QS}(n,n';t)$, as evident from Eq.~\ref{eq:EADMs}. From the Wigner 3j symbols in Eq.~\ref{eq:AKQS}, we determine allowed values of $K= 0, 1, 2$  and $S = 0, \pm1, \pm2$. $Q = 0$ for excitation by linearly polarized light, or $Q = 0, \pm1$ in general. We can further separate these into selection rules for coherences between excited states, and between the ground and excited state. For coherences between the ground ($n=1$) and excited state, $J_1 = 0$ and $J'_{n'} = 1$ are the only allowed excited-state angular momenta. This renders the first 3j symbol in Eq.~\ref{eq:AKQS} zero unless $K =1$. In other words, coherences between the ground and excited states are always oriented, and this orientation varies in time as $A^1_{0S}(1,n';t)$. This physically manifests as an oscillating dipole, as previously discussed. Similarly, the first 3j symbol restricts $K = 0, 2$ for coherences between excited states. In other words these coherences will always be aligned, but never oriented, with an alignment varying in time with $A^2_{0S}(n,n';t)$. Note that the MADM $A^0_{00}(n,n';t)$ is an isotropic moment and does not induce orientation or alignment (as discussed below, this is the reduced vibronic density matrix). This is the only moment allowed by Eq.~\ref{eq:AKQS} for the ground state population, while excited populations admit $K = 0$ and $K = 2$ moments. This essentially indicates that molecules in the excited state may be aligned but not oriented. Based on the Wigner-Ekart Theorem~\cite{zare1988,sakurai}, for $N$-photon excitation from an isotropic ground state, we can simply extend values of $K$ up to $2N$, with ground state - excited coherences only admitting the $K = N$ moment, and excited state coherences and populations only admitting even $K$. 
 
 These selection rules significantly limit the number of MADMs that need to be calculated to determine the density matrix. The computational burden can be further reduced by determining relationships between some of the MADMs. Since the Wigner Matrix Elements $D^K_{QS}(\mathbf{\Omega})$ are orthogonal~\cite{zare1988} the MADMs can be written as 
 
 \begin{equation}
     A^{K}_{QS}(n,n';t) = \frac{2K+1}{8\pi^2}\int d\mathbf{\Omega} D^{K}_{QS}(\mathbf{\Omega})\rho_{nn'}(\mathbf{\Omega},t). 
     \label{eq:MADMS}
 \end{equation}
 
 Using $\rho_{n'n}(\mathbf{\Omega},t) = \rho_{nn'}^*(\mathbf{\Omega},t)$ and $D^{K}_{-Q-S} = (-1)^{Q-S}D^{K*}_{QS}$, we find the relation
 
  \begin{equation}
     A^{K}_{-Q-S}(n',n;t) = (-1)^{Q-S} A^{K*}_{QS}(n,n';t).  
 \end{equation}
 
 This along with the selection rules limits the number MADMs that need to be calculated in any particular case. However, Eq.~\ref{eq:EADMs} indicates that the full rovibronic time-depdendent density matrix, $\rho_{\omega_n,\omega'_{n'}}(n,n';t)$, is needed to compute the MADMs. For most cases this is not computationally feasible~\cite{mayer1996}. As previously discussed, most computations of vibronic dynamics occur in the molecular frame, for a molecule at a fixed orientation. In the following subsection we discuss computation in this fixed-molecule approximation.
 
 \subsubsection{The fixed molecule approximation and the reduced vibronic density matrix}
 
 Calculations carried out in the molecular frame assume that the molecule is in an eigenstate of the orientation angles $\ket{\mathbf{\Omega}}$. We may expand this in the $\ket{JKM}$ basis,
 \begin{equation} \ket{\mathbf{\Omega}}=\sum_{JKM}\ket{JKM}\braket{JKM}{\mathbf{\Omega}},
 \end{equation}
 from which it is evident that an infinite superposition of $\ket{JKM}$ states is required to construct $\ket{\mathbf{\Omega}}$. This is not practically achievable. We note here that while quantum mechanics precludes the preparation of $\ket{\mathbf{\Omega}}$ states, it allows \emph{projection} of these onto a prepared state, $\left<\Omega|\psi\right>$, which constitutes a measurement of a quantity at fixed $\mathbf{\Omega}$. Coincidence detection of dissociative photoionization in the axial recoil approximation can pick a particular orientation from an underlying distribution, and is an example of such measurements~\cite{toffoli2007}. On the other hand, impulsive alignment experiments aim to prepare a state approximating $\ket{\mathbf{\Omega}}$ by laser-excitation of a broad superposition of $\ket{JKM}$ states. In general, such distributions are difficult to achieve for polyatomic molecules under laser-field free conditions~\cite{koch2019}. Furthermore, the Pauli Exclusion Principle obviates the existence of such approximate orientation eigenstates for certain molecules~\cite{grohmann2021}. As discussed in in section~\ref{sec:qtomography} below on measurable quantities and quantum tomography, the lab frame density matrix and the MADMs serve as an improved, direct comparison with experimental data. These may be computed using the fixed molecule approximation as well.
 
 This is achieved by recognizing that the matrix elements $\rho_{nn'}(\mathbf{\Omega},t)$ are in fact matrix elements of the density operator in the $\ket{\mathbf{\Omega} n}$ basis, diagonal in $\ket{\mathbf{\Omega}}$, $\bra{\mathbf{\Omega} n} \hat{\rho}(t) \ket{n' \mathbf{\Omega}}$. A derivation is presented in Appendix \ref{appendix:madms}. We note that there are also density matrix elements which are off-diagonal in $\ket{\mathbf{\Omega}}$. These cannot be written in terms of the MADMs but are evidently not needed for the determination of vibronic dynamics (cf. Eq~\ref{eq:PintermB}). Given this, the standard vibronic calculations can be carried out to determine the vibronic (or electronic, depending on the problem) density matrix $\rho_{n,n'}(t)$ at different orientation angles $\mathbf{\Omega}$ with respect to the exciting pump pulse to yield $\rho_{n,n'}(\mathbf{\Omega}, t)$. While this is still an increased computational burden over a single  molecular frame calculation, since only moments up to $K = 2$ are excited by one photon relatively low orientation-angle resolution will suffice. The integral in Eq.~\ref{eq:MADMS} can then be carried out to compute the MADMs to serve as a basis for comparison with experiment (cf section~\ref{sec:qtomography}).  
 
 With the identification $\rho_{nn'}(\mathbf{\Omega},t) = \bra{\mathbf{\Omega} n} \hat{\rho}(t) \ket{n' \mathbf{\Omega}}$, we have the normalization condition $Tr\{\hat{\rho}\} = 1$, or 
 \begin{equation}
     \sum_{n}\int{\rho_{nn}(\mathbf{\Omega},t)d\mathbf{\Omega}}=1.
 \end{equation}
 This along with Eq.~\ref{eq:ADMOrigin} and the integral of the Wigner Matrix Elements~\cite{zare1988} provides the following normalization for the MADMs,
 \begin{equation}
     \sum_{n}A^0_{00}(n,n;t) = \frac{1}{8\pi^2}.
 \end{equation}
 Additionally, integrating Eq.~\ref{eq:ADMOrigin} and~\ref{eq:EADMOrigin} over all orientations provides the experimentally accessible reduced vibronic density matrix,
 \begin{equation}
     \rho_{nn'}(t) = 8\pi^2 A^0_{00}(n,n';t).
     \label{eq:reddensmat}
 \end{equation}
 Thus, the orientation-averaged reduced vibronic density matrix includes only coherences that have a non-zero isotropic MADM in the lab frame ($A^0_{00}(n,n';t)\neq0$). A direct consequence of this result and the selections rules from the previous sub-section is that the coherences between the ground and excited state are \emph{not} included in the reduced vibronic density matrix for single photon excitation from an isotropic ensemble, since such coherences only have a $K = 1$ MADM. Any experiment designed to specifically target the reduced vibronic density matrix will not be sensitive to these coherences, if the experiment is initiated from an isotropic ensemble. On the another hand, as discussed in~\cite{makhija2020} and section~\ref{sec:qtomography}, experiments that measure the anisotropy in the distribution of a particle scattered off the excited molecule will potentially be directly sensitive to a number of MADMs. Thus the orientation averaged reduced vibronic density matrix does not accurately represent the information available from such experiments. 
 
 While the fixed molecule approximation is useful, particularly for expensive vibronic calculations, a calculation which treats the rotational states explicitly is preferable yielding the vibronic dynamics at any orientation angle from Eq.~\ref{eq:PintermB}, and equivalently the full lab frame molecular probability distribution. In particular, this approximate method fails in cases in which rotational and electronic degrees of freedom are coupled~\cite{mayer1996,makhija2020}. In the following section we present a numerical calculation of only electronic dynamics within the rigid rotor and BO approximations that demonstrates the concepts presented thus far, and verifies some of the theoretical results.
 
 \section{A Numerical Example}
A rigorous calculation of $\rho_{\alpha \alpha' v_\alpha v'_{\alpha'}}(\mathbf{\Omega},t)$ in Eq.~\ref{eq:PintermB} demands considering all molecular degrees of freedom. Since this is impractical in most scenarios, here we neglect vibrational degrees of freedom in order to demonstrate some of the general features of $\rho_{\alpha \alpha' v_\alpha v'_{\alpha'}}(\mathbf{\Omega},t)$ discussed above. We consider a superposition of molecular eigenstates which can be excited by a broad band laser pulse:
\begin{equation}
    \ket{\Psi(t)}=\sum_{J_\alpha M_\alpha \tau_\alpha  \alpha} a^\alpha_{J_\alpha M_\alpha \tau_\alpha}\exp\left(-iE^\alpha_{J_\alpha M_\alpha \tau_\alpha }t\right)\ket{\alpha}\ket{J_\alpha M_\alpha \tau_\alpha }.
    \label{eq:Psiket}
\end{equation}
Here we assume that the BO states $\ket{\alpha}\ket{J_n M_\alpha \tau_\alpha }$ are eigenstates of the Hamiltonian with energies $E^\alpha_{J_\alpha M_\alpha \tau_\alpha } = E_\alpha + E_{J_\alpha M_\alpha \tau_\alpha }$, $E_\alpha$ and $ E_{J_\alpha M_\alpha \tau_\alpha }$ being the electronic and rotational energies, respectively. This level of calculation is useful in situations where only electronic dynamics are of interest, such as in attosecond charge migration. That is, when considering only very short times after the pulse, during which the nuclei can be satisfactorily approximated as frozen. The states $\ket{J_\alpha M_\alpha \tau_\alpha }$ are eigenstates of the asymmetric top Hamiltonian~\cite{zare1988}, and can be determined in the symmetric top basis,
\begin{equation}
    \ket{J_\alpha M_\alpha \tau_\alpha}=\sum_{K_\alpha} C^{J_\alpha M_\alpha \tau_\alpha}_{K_\alpha} \ket{J_\alpha K_\alpha M_\alpha}
    \label{eq:JtM}
\end{equation}
given a set of rotational constants $A = 1/2I_{aa}$, $B = 1/2I_{bb}$ and $C = 1/2I_{aa}$, where $I_{jj}$ are diagonal components of the moment of inertia tensor with $I_{cc}>I_{bb}>I_{aa}$. Since we assume a frozen geometry, we take these to be the same for all electronic states $\ket{\alpha}$ in the superposition. Doing so gives each electronic state the same set of rotational eigenstates. We also note here that, unlike $K_\alpha$, $\tau_\alpha$ is not an angular momentum projection on the molecular frame. This projection is not a conserved quantity for an asymmetric top. Rather, $\tau_\alpha$ is an index used to keep track of the set of energy eignestates with the same $J_\alpha$ and $M_\alpha$~\cite{zare1988}.

\begin{figure}
    \centering
    \includegraphics[scale=0.6,trim=4 4 4 4,clip]{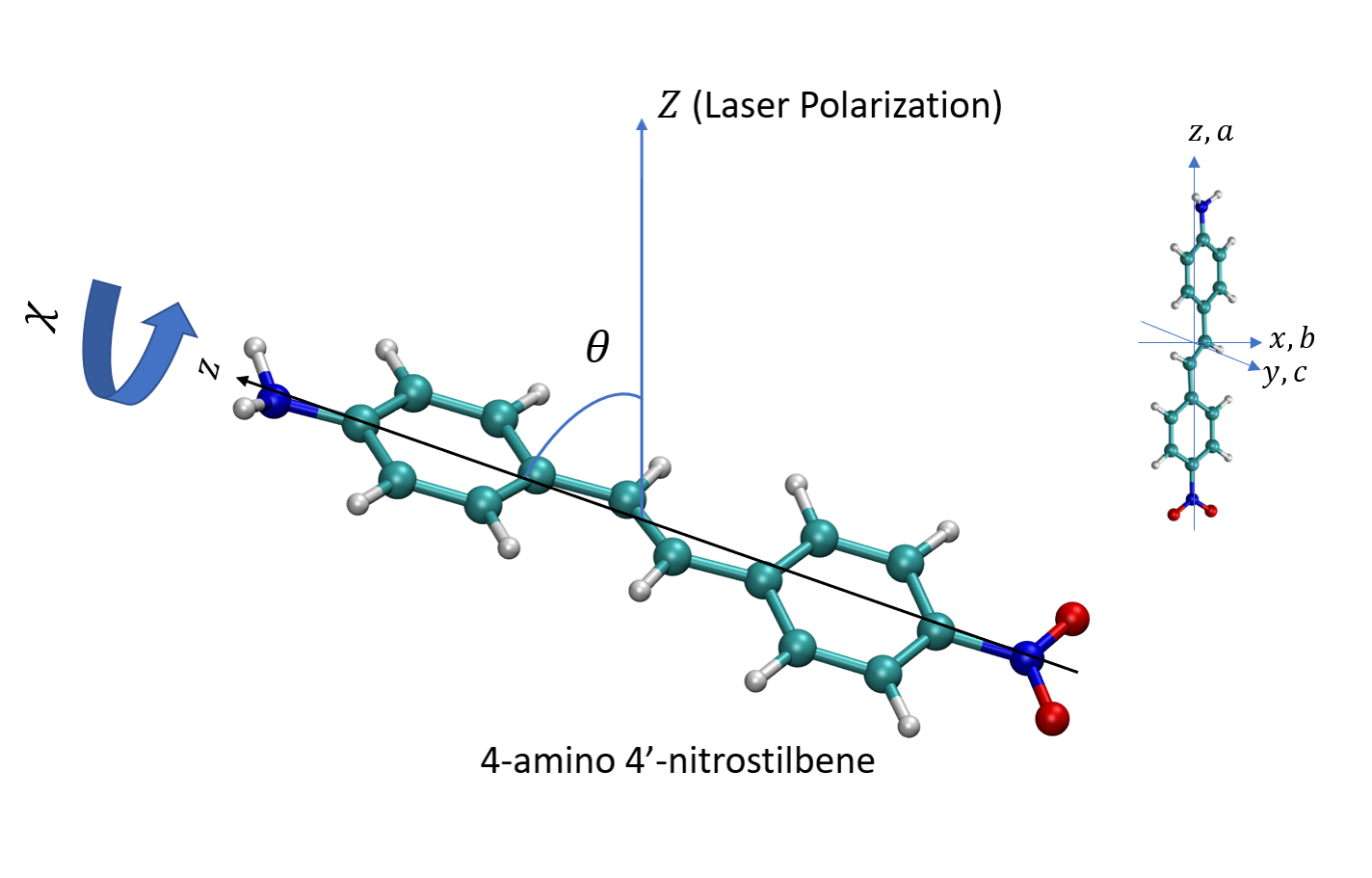}
    \caption{The ground state geometry of 4-amino-4$'$-nitrostilbene, with the Euler angles $\theta$ and $\chi$ labeled with respect to the lab frame $Z$-axis.The molecular frame coordinate system is also shown in the inset.}
    \label{fig:geom}
\end{figure}

The system chosen here is 4-amino-4$'$-nitrostilbene excited to the manifold of three closely-spaced bright $\pi\pi^{*}$ states centred around 4 eV. The required electronic energies $E_{\alpha}$ and transition dipole moments $\mu_{0\alpha,i}=\bra{\alpha}\hat{\mu}_i\ket{0}$ were computed at the pruned combined density functional theory and multireference configuration interaction (p-DFT/MRCI)\cite{neville2021} in conjunction with the def2-TZVP basis. These values are shown in Table~\ref{tab:Energy/States}. The inertia tensor $I_{jj}$ of the ground electronic state was calculated at the DFT level of theory using the TZVP basis and B3LYP exchange correlation functional. The resulting principal moments of inertia are  $I_{aa}=821.87$~ amu$\cdot$Bohr$^2$, $I_{bb}= 15275.20$~amu$\cdot$Bohr$^2$ and  $I_{cc}=16079.53$~amu$\cdot$Bohr$^2$, and the $z$-axis is identified with the $a$-axis, the $x$-axis with the $b$-axis and the $y$-axis with the $c$-axis. The molecular geometry with the molecular frame axes labelled is depicted in the inset in Fig.~\ref{fig:geom}. The inertia tensor is used to compute the rotational energies $E_{J_\alpha M_\alpha \tau_\alpha}$, which are considered to be the same for all electronic states. These are determined for the asymmetric top Hamiltonian following the procedure outlined in ~\cite{zare1988}.   

\begin{table}[h]
    \centering
    \begin{tabular}{|c|c|c|c|c|c|}
    \hline
         $\alpha$ & State & $E_\alpha$ (eV) & $\mu_{0\alpha,z}$ (au) & $\mu_{0\alpha,x}$ (au) &$\mu_{0\alpha,y}$ (au)  \\
         \hline
         2 & $S_{4}(\pi\pi^{*})$ & 3.87 & 1.07552 & 0.10603 & -0.01137 \\
         \hline
         3 & $S_{5}(\pi\pi^{*})$ & 4.04 & 0.53848 & 0.18773 & 0.01271 \\
         \hline
         4 & $S_{6}(\pi\pi^{*})$ & 4.08 & -0.84664 & 0.32543 & -0.01292\\
         \hline
    \end{tabular}
    \caption{Excitation energies, $E_{\alpha}$, and transition dipole matrix elements, $\mu_{0\alpha,i} = \bra{\alpha}\hat{\mu}_i\ket{0}$, for the states included in the 4-amino-4$'$-nitrostilbene calculations. All quantities were calculated at the p-DFT/MRCI level of theory with the def2-TZVP basis set.}
    \label{tab:Energy/States}
\end{table}

We use first order perturbation theory, implying a one-photon process, in a linearly polarized electric field to determine the amplitudes $a^\alpha_{J_\alpha M_\alpha \tau_\alpha}$ in Eq.~\ref{eq:Psiket},
\begin{equation}
    a^{\alpha}_{J_\alpha M_\alpha \tau_\alpha} \propto \bra{J_\alpha M_\alpha \tau_\alpha }\bra{\alpha}\mu_Z \ket{0}\ket{J_{in} M_{in} \tau_{in}}
    \label{eq:a1}
\end{equation}
where the overall multiplicative factor is determined by the bandwidth of the pulse. We ignore this factor, hence assuming a flat-top bandwidth across the set of states in table~\ref{tab:Energy/States}.  In this limit, the excitation pulse is much shorter than the timescale of electronic quantum dynamics which, in turn, is determined by the spacing between electronic states involved. Eq.~\ref{eq:a1} implies that the system starts in the ground state $\ket{J_{in} M_{in} \tau_{in}} \ket{0}$ and is excited to the state $\ket{J_\alpha M_\alpha \tau_\alpha }\ket{\alpha}$.

The transition dipole moments in Table~\ref{tab:Energy/States} are in the molecular frame while $\mu_Z$ in Eq.~\ref{eq:a1} is in the lab frame. In order to transform Eq.~\ref{eq:a1} into the molecular frame, we convert $\mu_Z$ into a spherical tensor~\cite{zare1988}. In the spherical basis, $\mu_Z=\mu^1_0$ and can be expressed in terms of the molecular frame spherical dipole moments $\mu^1_q$:   
\begin{equation}
    [\mu^1_0]^{LF}=\sum_{q}D^{1*}_{0q}(\mathbf{\Omega})\mu^1_{q}
    \label{eq:muEz}
\end{equation}
Here, $\mu^1_{q}$ are the spherical components of the dipole moment in the molecular frame with values of $q$ ranging from -1 to 1. The irreducible spherical components of the molecular frame dipole moment are as follows ~\cite{zare1988}:
\begin{equation}
    \mu^{1}_{-1}= \frac{1}{\sqrt{2}}(\mu_{x} - i\mu_{y}) 
\end{equation}
\begin{equation}
    \mu^{1}_{1}=-\frac{1}{\sqrt{2}}(\mu_{x} + i\mu_{y})
\end{equation}
\begin{equation}
    \mu^{1}_{0}= \mu_{z}.
\end{equation}
Using Eq.~\ref{eq:muEz} in Eq.~\ref{eq:a1} gives the following expression for $a^{\alpha}_{J_\alpha M_\alpha\tau_\alpha}$. 
\begin{equation}
    a^{\alpha}_{J_\alpha M_\alpha\tau_\alpha}=\sum_q \bra{J_\alpha M_\alpha \tau_\alpha }D^{1*}_{0q}\ket{J_{in}M_{in}\tau_{in}}\bra{\alpha}\mu^1_q\ket{0},
\end{equation}
which, with the use of Eq.~\ref{eq:JtM}, can be expressed as a sum over $q, K_{in}$ and $K_{\alpha}$:
\begin{equation}
    a^{\alpha}_{J_\alpha M_\alpha\tau_\alpha}=\sum_q \sum_{K_{in}K_{\alpha}}C^{J_{in} M_{in}\tau_{in}}_{K_{in}}C^{J_{\alpha} M_{n} \tau_{n}*}_{K_{\alpha}}\bra{J_\alpha K_\alpha M_\alpha}D^{1*}_{0q}\ket{J_{in}K_{in}M_{in}}\bra{\alpha}\mu^1_q\ket{0}.
\end{equation}
The set of dipole matrix elements, calculated at the p-DFT/MRCI/def2-TZVP level of theory, are given in table~\ref{tab:Energy/States}. The matrix elements $\bra{J_\alpha K_\alpha M_\alpha}D^{1*}_{0q}\ket{J_{in}K_{in}M_{in}}$ are straightforward to determine analytically in terms of Wigner 3j symbols~\cite{zare1988}.This finally gives
\begin{equation}
\begin{split}
    a^{\alpha}_{J_\alpha M_\alpha\tau_\alpha}=\sum_q \sum_{K_{in}K_{\alpha}}C^{J_{in} M_{ in} \tau_{in}}_{K_{in}}C^{J_{\alpha} M_{n} \tau_{\alpha}*}_{K_{\alpha}}\sqrt{(2J_\alpha+1)(2J_{in}+1)}(-1)^{q}(-1)^{K_{in}-M_{in}} \times \\ \begin{pmatrix} J_{in}& 1 &J_\alpha \\ -M_{in} & 0 & M_\alpha \end{pmatrix}\begin{pmatrix} J_{in}& 1 &J_\alpha \\ -K_{in} & -q & K_\alpha \end{pmatrix}\bra{\alpha}\mu^1_q\ket{0}
\end{split}
\end{equation}

In order to calculate the MADMs Eq.~\ref{eq:Psiket} must be rewritten in the $\ket{J_\alpha K_\alpha M_\alpha}$ basis:
\begin{equation}
    \ket{\Psi(t)}=\sum_{J_\alpha K_\alpha M_\alpha \alpha}\left(\sum_{\tau_\alpha} a^\alpha_{J_\alpha M_\alpha \tau_\alpha }C^{J_\alpha M_\alpha\tau_\alpha}_{K_\alpha} \exp\left(-iE^\alpha_{J_\alpha M_\alpha \tau_\alpha }t\right)\right) \ket{\alpha} \ket{J_\alpha K_\alpha M_\alpha}
    \label{eq:psi}
\end{equation}
The coefficient in parenthesis provides the amplitude in the $\ket{J_\alpha K_\alpha M_\alpha}$ basis, 
\begin{equation}
    x_{J_\alpha K_\alpha M_\alpha}(t)=\sum_{\tau_\alpha} a^\alpha_{J_\alpha M_\alpha \tau_\alpha}C^{J_\alpha M_\alpha\tau_\alpha}_{K} \exp(-iE^\alpha_{J_\alpha M_\alpha \tau_\alpha }t)
\end{equation}
leading to the density matrix elements:
\begin{equation}
     x_{J_\alpha K_n M_\alpha}(t)x^*_{J'_{\alpha'} K'_{\alpha'} M_\alpha}(t)=\rho_{{\omega}_\alpha {\omega}'_{\alpha'}}(\alpha,\alpha';t)
\end{equation}
The MADMs can then be determined using Eq.~\ref{eq:EADMs}. In an experiment the initial condition is typically a thermal distribution of rotational states in the ground vibronic state. To simulate this we repeat the calculation for a number of initial rotational states $\ket{J_{in} M_{in} \tau_{in}}$, and Boltzmann average over the resulting set of MADMs. The necessary number of initial states can be determined by summing the Boltzmann probability distribution over energy. Once the statistical probability reaches $.99$, the sum is terminated, the last state in the sum being the highest energy initial state considered. Thermally averaged MADMs at a temperature of 0.1~K, corresponding to 84 initial rotational states, are used in Eq.~\ref{eq:ADMOrigin} and Eq.~\ref{eq:EADMOrigin} to calculate the lab frame density matrix elements. 

\begin{figure}
    \centering
    \includegraphics[scale=0.45,trim=4 4 4 4,clip]{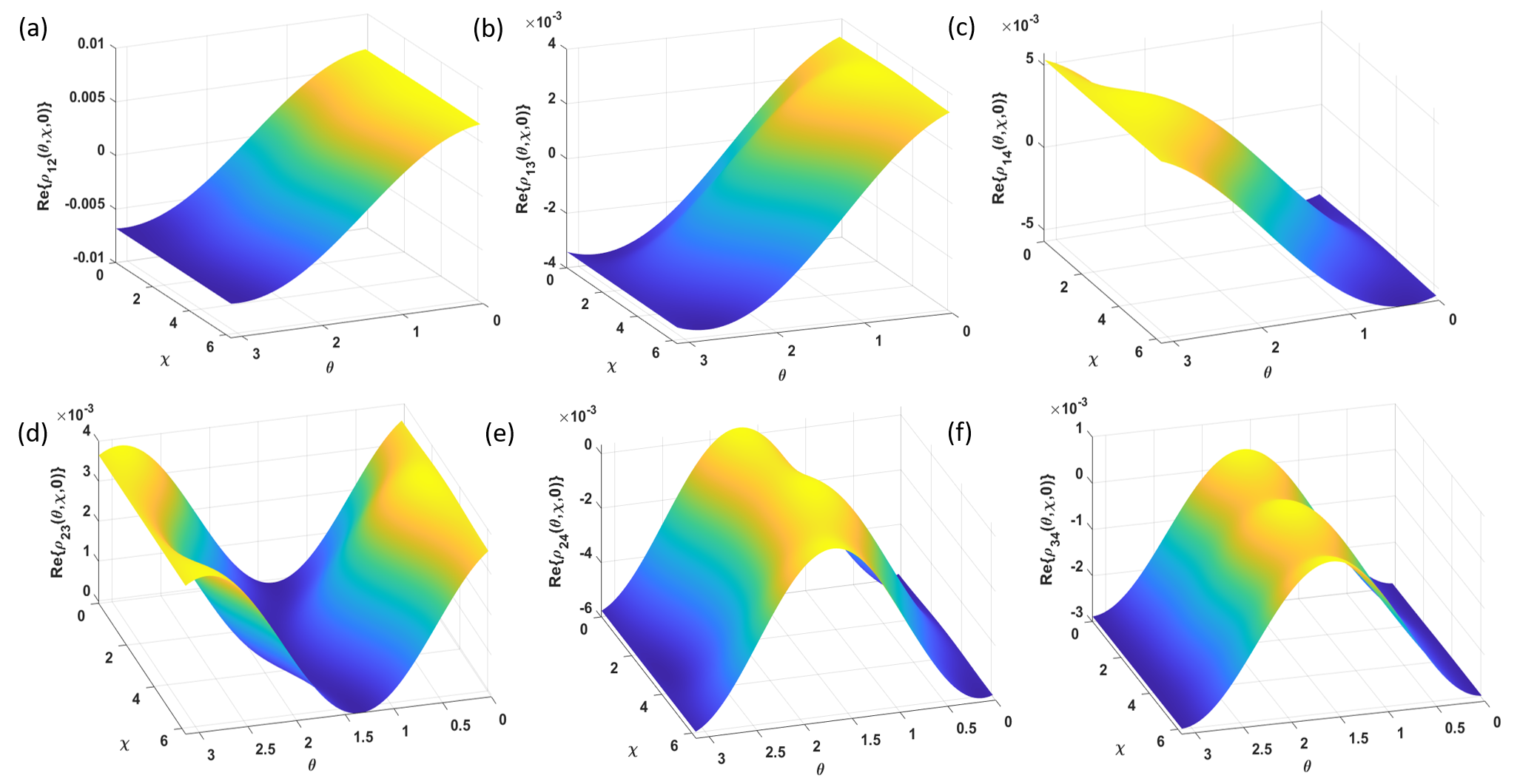}
    \caption{(a)-(c) Initial orientation-dependent off-diagonal density matrix elements between the ground and excited states, $\rho_{12}(\theta,\chi,0), \rho_{13}(\theta,\chi,0),\rho_{14}(\theta,\chi,0)$. (d)-(f) Initial orientation-dependent off-diagonal density matrix elements between excited states, $\rho_{23}(\theta,\chi,0), \rho_{24}(\theta,\chi,0),\rho_{34}(\theta,\chi,0)$.}
    \label{fig:surfs}
\end{figure}

\begin{figure}
    \centering
    \includegraphics[scale=.45,trim=4 4 4 4,clip]{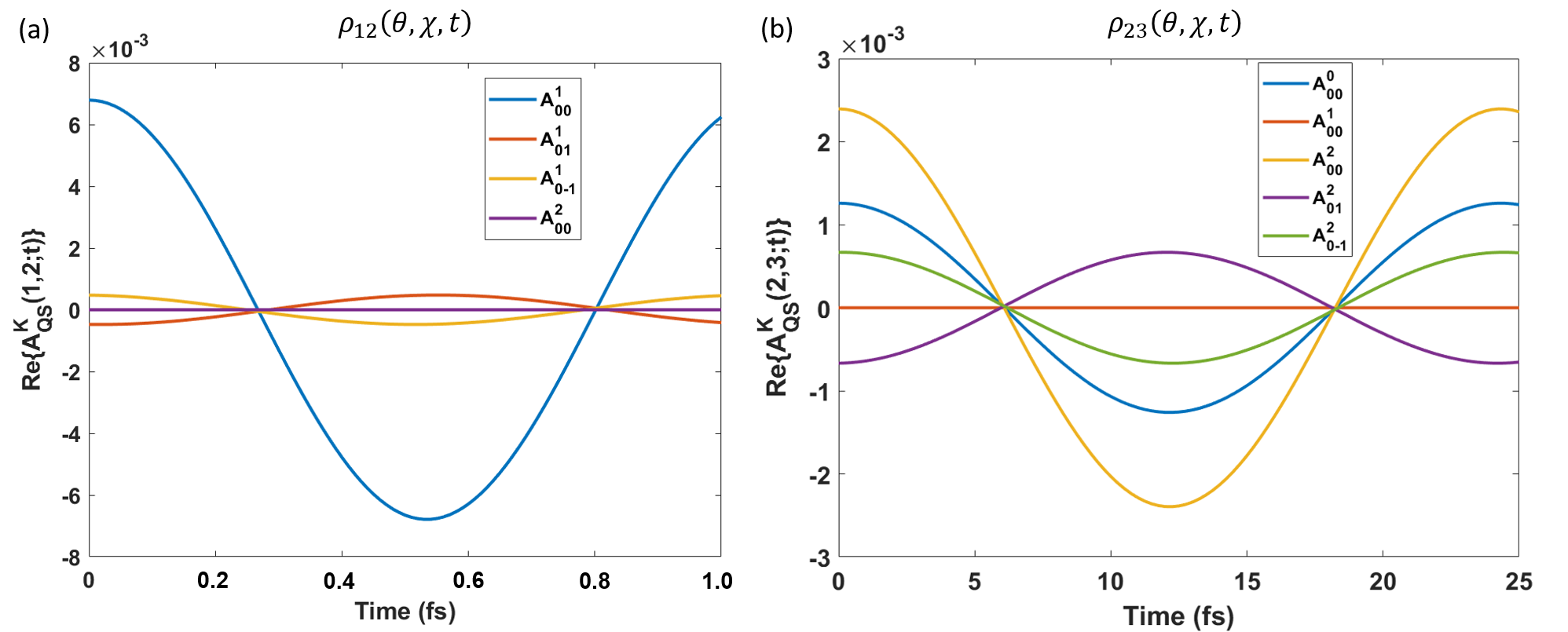}
    \caption{(a) MADMs contributing to the ground state-excited state density matrix element $\rho_{12}(\theta,\chi,t)$. The MADM $A^{2}_{00}(1,2;t)$ is always zero in agreement with the selection rules detailed in the previous section, as are all the MADMs $A^2_{0S}(1,2;t)$ and $A^0_{00}(1,2;t)$  (not shown).(b) Some of the MADMs contributing to the excited states density matrix element $\rho_{23}(\theta,\chi,t)$. The MADM $A^{1}_{00}(1,2;t)$ is always zero in agreement with the selection rules detailed in the previous section, as are all the MADMs $A^1_{0S}(1,2;t)$ (not shown). For all density matrix elements MADMs with $K>2$ are zero in agreement with the selection rules.}
    \label{fig:MADMs}
\end{figure}

Fig.~\ref{fig:geom} shows the geometry of 4-amino-4$'$-nitrostilbene as well as the relevant orientation angles $\mathbf{\Omega}=\{\theta,\chi\}$ with respect to the laser polarization. All quantities are independent of the azimuthal angle $\phi$ due to cylindrical symmetry. In order to illustrate the analytical results derived above, the calculated $\rho_{\alpha \alpha' }(\mathbf{\Omega},t)$ are plotted at $t = 0$ as a function of $\theta$ and $\chi$ in Fig.~\ref{fig:surfs}. We show only the real parts, as the density matrix elements are all real at $t=0$. The top row, Figs.~\ref{fig:surfs} (a) to (c), show coherences between the ground and excited states. Evidently, these coherences are oriented - their values for molecules oriented at $\theta, \chi$ and $\theta + \pi, \chi + \pi$ are different thus breaking inversion symmetry in the lab frame. This is anticipated by the selection rule $K = 1$ for one-photon excited states. Conversely, the coherences in the bottom row are aligned and do not break inversion symmetry in the lab frame, in accordance with the selection rule $K = 0, 2$. Fig.~\ref{fig:MADMs}(a) shows the relevant MADMs for the coherence between the ground and first excited state (labelled 2 in Table~\ref{tab:Energy/States}). Only the MADMs $A^{1}_{0S}(1,2;t)$ are non-zero, representing the oscillating dipole induced by the excitation. Note that this decays over time due to spontaneous emission, not included in our calculation. Similarly, in Fig.~\ref{fig:MADMs}(b) we see that the MADMs with $K = 1$ are zero for the coherence between the excited states labeled $2$ and $3$, and MADMs $A^{0}_{00}(2,3;t)$ and $A^{2}_{0S}(2,3;t)$ are non-zero.   

\begin{figure}
    \centering
    \includegraphics[scale=.46,trim=4 4 4 4,clip]{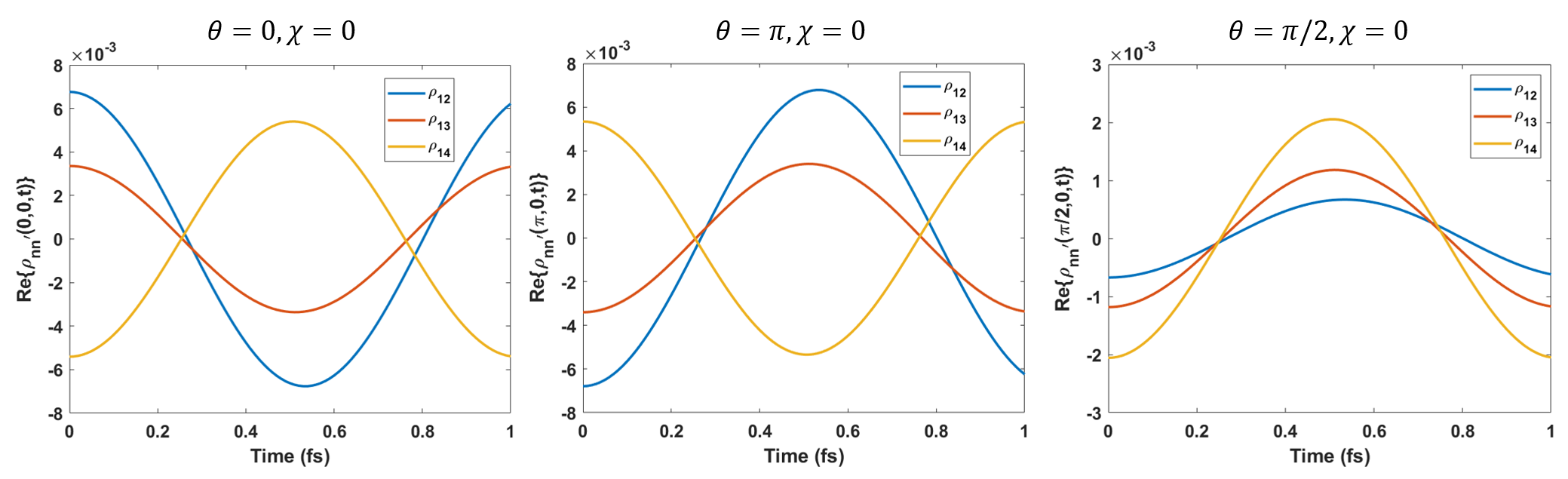}
    \caption{ground state-excited state density matrix elements $\rho_{12}(\theta,\chi,t)$, $\rho_{13}(\theta,\chi,t)$ and $\rho_{14}(\theta,\chi,t)$ at orientations \{$\theta = 0,\chi = 0$\}, \{$\theta = \pi,\chi = 0$\} and \{$\theta = \pi/2,\chi = 0$\}, respectively.}
    \label{fig:dynamics}
\end{figure}

To elucidate the orientation dependent electronic dynamics, we plot the time-depdendent coherences between the ground state and excited states in Fig.~\ref{fig:dynamics}. Note that we are treating these as eigenstates, rendering the populations constant. Further, Fig.~\ref{fig:MADMs}(b) indicates that in this case coherences between excited states evolve on a much slower time scale, as a result of the much smaller energy spacing. While our calculations within the rigid body approximation do not include vibration, we note that including the vibrational degrees of freedom will drive the components of the wave packet in different electronic states apart, eventually destroying the coherence. In general, the timescale for decoherence will be sensitive to the differential topology of the potential energy surfaces involved. Based on such consideration, recent studies indicate that the coherences could decay significantly within a range of time windows, anywhere between 2 and 10 fs~\cite{arnold2017,arnold2018,vacher2017,despre2018}. We therefore do not consider the slow excited state - excited state coherences, like that in Fig.~\ref{fig:MADMs}(b), any further. In Fig.~\ref{fig:dynamics}, we focus on the early time dynamics, during which the excited state-excited state coherences are essentially constant. Comparing the phase relation of the set of coherences for a molecule oriented at $\theta = 0$, $\chi = 0$ to one at $\theta = \pi$, $\chi = 0$, we see that we can expect very similar dynamics in both cases, but with a relative time delay of approximately $0.5$~fs between the two since each coherence is shifted in phase by $\pi$. On the other hand, we can expect very different dynamics for a molecule at $\theta = \pi/2$, $\chi = 0$, since the phase relation between the coherences is completely different. This illustrates the general point that the molecular frame dynamics occurring in an experiment will vary for molecules in different orientations. While we have held the angle $\chi$ fixed here, Fig.~\ref{fig:surfs} shows that the coherences also vary with $\chi$ indicating that the molecular frame dynamics will vary accordingly. 
\begin{figure}
    \centering
    \includegraphics[scale=.15]{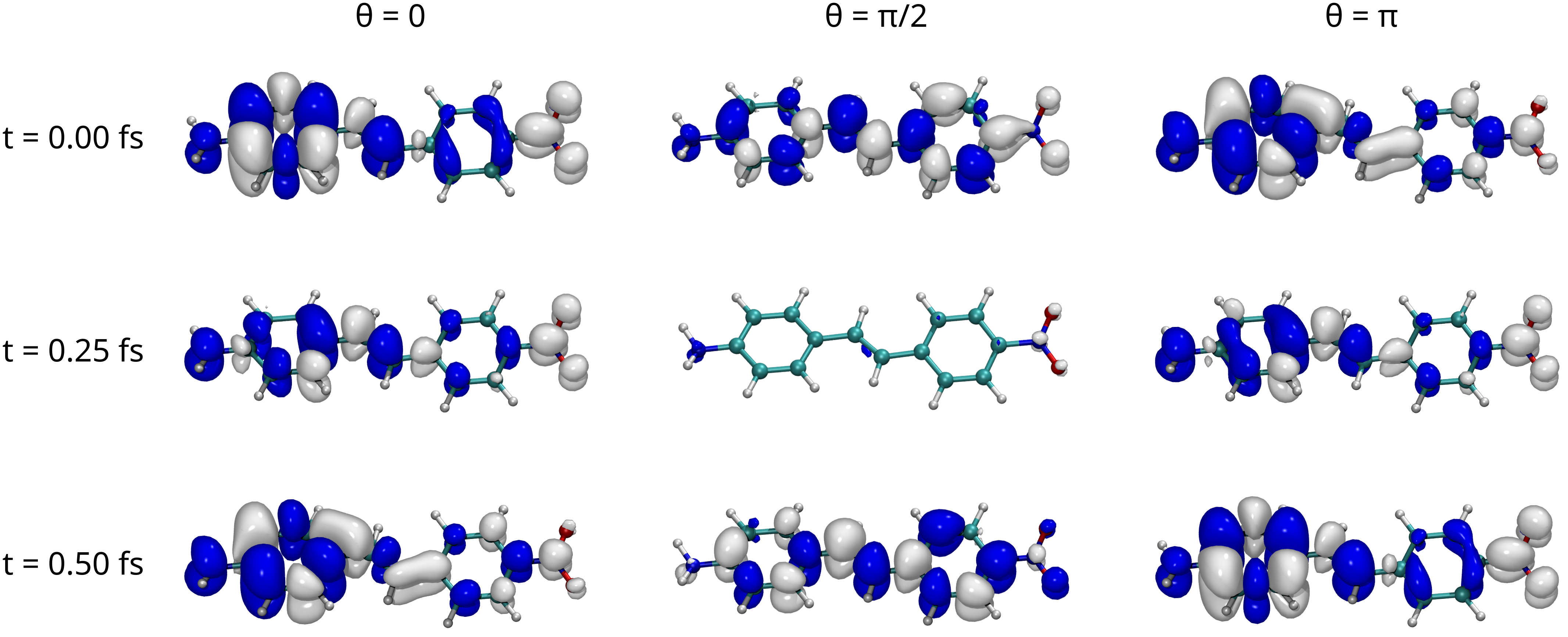}
    \caption{Time evolution of the one-electron difference density $\Delta p(\boldsymbol{r}_{1}, \mathbf{\Omega}, t)$ for various orientation angles $\mathbf{\Omega} = \{ \theta, \chi=0\}$ and times $t$. The regions colored white correspond to a net-positive change, i.e., accumulation, of electron density relative to the ground electronic state. Those colored blue correspond to net-negative change, i.e., depletion, of electron density.}
    \label{fig:electrondensity}
\end{figure}

In Fig.~\ref{fig:electrondensity}, we show the time evolution of the one-electron difference density $\Delta p(\boldsymbol{r}_{1}, \mathbf{\Omega}, t)$ for various orientation angles $\mathbf{\Omega} = \{ \theta, \chi=0 \}$. $\Delta p(\boldsymbol{r}_{1}, \mathbf{\Omega}, t)$ is here defined as the difference of the wave packet and ground state electronic probability distributions integrated over all but one electronic degree of freedom,

\begin{equation}\label{eq:diffdens}
\Delta p(\boldsymbol{r}_{1}, \mathbf{\Omega}, t) = N \int d\boldsymbol{r}_{2} \cdots d\boldsymbol{r}_{N} \left[ P(\boldsymbol{r}_{1},\dots,\boldsymbol{r}_{N}, \mathbf{\Omega}, t) - P_{0}(\boldsymbol{r}_{1},\dots,\boldsymbol{r}_{N}) \right].
\end{equation}

\noindent
The details of these calculations are given in Appendix \ref{appendix:diffdens}. The expected phase difference between molecules oriented at $\theta = 0$ and $\theta = \pi$ is evident. Additionally, molecules at $\theta = \pi/2$ exhibit entirely different dynamics. Furthermore, these dynamics are driven entirely by the set of ground state-excited state coherences shown in Fig.~\ref{fig:dynamics}, which are in turn determined solely by the the MADMs $A^1_{0S}(1,\alpha';t)$. Since the orientation-averaged reduced vibronic density matrix is $8\pi^2A^0_{00}(\alpha,\alpha';t)$ (cf Eq.~\ref{eq:reddensmat}), these dynamics are completely invisible to an experiment designed to specifically measure the reduced vibronic density matrix elements but may, in general, be accessible experimentally. In fact, the variety of dynamics evident in Fig.~\ref{fig:electrondensity} indicates that a great deal of information is potentially contained in a lab frame experiment on un-oriented molecules, but this information is difficult to extract. The next section discusses the possibility of extracting this information, and the limitations associated with such an extraction.

\section{Measurements and Quantum Tomography}
\label{sec:qtomography}
The MADMs shown in Eq.~\ref{eq:ADMOrigin} and Eq.~\ref{eq:EADMOrigin} to be the time-depdendent components of the lab frame density matrix provide a direct connection between the density matrix and measurements. Determination of this density matrix from measurement then allows the expectation value of any \emph{molecular frame} operator $\hat{O}$ to be determined for a molecule at any orientation angle,
\begin{equation}
    \left < \hat{O} \right > (\mathbf{\Omega},t) = \sum_{nm}\rho_{nm}(\mathbf{\Omega},t) \bra{m}\hat{O}\ket{n}.
\end{equation}
Since $\hat{O}$ is defined in the molecular frame, it is exclusively a function of vibronic coordinates and momenta. Therefore it acts as the unit operator on $\ket{\mathbf{\Omega}}$, and density matrix elements that are off-diagonal in $\ket{\mathbf{\Omega}}$ are not needed to calculate its expectation value. Note that these are needed to determine expectation values of lab frame operators, such as the total angular momentum. As previously discussed, these cannot be written in terms of the MADMs. Determination of the lab frame density matrix also provides the full molecular probability density through Eq.~\ref{eq:PintermB}, and therefore the time varying molecular frame vibronic probability density for a molecule at any orientation angle. We can therefore refer to the determination of all the relevant MADMs as a molecular frame quantum tomography. To perform this tomography we must consider the link between specific measurements and the MADMs.
In particular, time resolved measurements of photoelectron angular distributions from a molecular wavepacket contain the MADMs. This has been shown theoretically~\cite{underwood2008,underwood2005,underwood2000} and recently demonstrated experimentally for an electronic wavepacket in the $B^{ 1}E''$ state of ammonia~\cite{makhija2020}. This relationship has previously been exploited to determine molecular frame photoelectron angular distributions using a rotational wavepacket in the ground state of isolated molecules~\cite{gregory2021,marceau2017}. In a similar study, the angular distributions of diffracted electron and x-ray pulses from a ground state rotational wavepacket were recently shown to contain the ADMs, and a method to extract the molecular geometry using this relationship for electron diffraction was numerically demonstrated~\cite{hegazy2022}. Recent ultrafast X-ray diffraction measurements from an excited state wavepacket imply a direct connection between the X-ray angular distribution and the MADMs~\cite{natan2021}, though this has yet to be shown analytically. A direct connection, such as that now known to exist for time resolved photoelectron angular distributions, provides a potential route to determining all relevant lab frame density matrix elements from experiment constituting a molecular frame quantum tomography. We discuss this possibility in more detail below.

Time resolved photoelectron angular distributions are typically analyzed by multipole expansion as follows,
\begin{equation}
\sigma(\epsilon,t,\theta_{e},\phi_{e})=\sum_{LM}\beta_{LM}(\epsilon,t)Y^{L}_{M}(\theta_{e},\phi_{e}),
\label{LFPAD}
\end{equation}
where $\epsilon$ is the kinetic energy of the electron and $\theta_{e}$ and $\phi_{e}$ its polar and azimuthal ejection angles. The time-depdendent anisotropy parameters $\beta_{LM}(\epsilon,t)$ can then be expressed in terms of the MADMs~\cite{underwood2008,hockett2018QMP1,hockett2018QMP2,makhija2020,gregory2021},
\begin{align}
\beta_{LM}(\epsilon,t)=\sum_{nn'}\sum_{KQS}C^{LM}_{KQS}(n,n';\epsilon) A^{K}_{QS}(n,n';t)
\label{eq:betaLF}\\
C^{LM}_{KQS}(n,n';\epsilon)=\sum_{\zeta\zeta'}D^{n}_{\zeta}(\epsilon)D^{n'*}_{\zeta'}(\epsilon)\Gamma^{\zeta\zeta'LM}_{KQS}.
\label{eq:coeffLF}
\end{align}
Eq.~\ref{eq:betaLF} separates the excited state dynamics - fully characterized by the MADMs - from the photoionization dynamics characterized by the coefficients $C^{LM}_{KQS}(n,n';\epsilon)$. These coefficients contain the ionization dipole matrix elements $D^{n}_{\zeta}(\epsilon)$ and $D^{n'*}_{\zeta'}(\epsilon)$ between the pair of bound states $n$ and $n'$ into the set of continuum channels $\zeta$ and $\zeta'$ accessible from each of the bound states. In other words, $C^{LM}_{KQS}(n,n';\epsilon)$ is zero unless both bound states $n$ and $n'$ ionize to the same final state producing an electron of energy $\epsilon$. The analytical coupling factor $\Gamma^{\zeta\zeta'LM}_{KQS}$ ensures that total angular momentum is conserved during the photoionization process. It has been previously studied in detail~\cite{gregory2021}, and we need not discuss it further here. Since a set of $\beta_{LM}(\epsilon,t)$ are typically measured in a time resolved experiment, if the $C^{LM}_{KQS}(n,n';\epsilon)$ are known the MADMs can potentially be determined from Eq.~\ref{eq:betaLF}, which are a set of linear equations in the MADMs at each time. Eq.~\ref{eq:ADMOrigin} and Eq.~\ref{eq:EADMOrigin} then provide the lab frame density matrix. 

A number of issues need to be carefully considered here. Firstly, experimental determination of the \\$C^{LM}_{KQS}(n,n';\epsilon)$, while possible is highly non-trivial. This requires a so-called complete experiment - determination of the ionization dipole matrix elements (typically in the partial wave basis) -  accomplished for a only a handful of molecules~\cite{reid1992,motoki2002,lebech2003,cherepkov2005,teramoto2007,hockett20071,hockett20072,hockett2009,marceau2017}. While a number of methods have been applied to this end, it has been most reliably accomplished for an excited state of a polyatomic molecule by rotationally resolved, resonance enhanced multiphoton ionization (REMPI) through the excited state of interest~\cite{hockett20071,hockett20072,hockett2009}. Even in cases where the ionization dipole matrix elements are known, an additional challenge is posed by Koopman's Correlations to the continuum~\cite{underwood2008}. Often, a single bound state is correlated with a single ionic state. While these correlations have proved invaluable for tracking populations of electronic states~\cite{blanchet1999,underwood2008}, they render the $C^{LM}_{KQS}(n,n';\epsilon)$ zero when $n\neq n'$ removing off-diagonal MADMs, and therefore any direct information about coherences from the measurement. Finally, in most cases there will not be a sufficient number of equations to uniquely determine all relevant MADMs needed to construct the density matrix. For one-photon excitation by a linearly polarized pulse followed by one-photon ionization by an elliptically polarized pulse $\beta_{LM}$ with $L = 0,2$ and $4$ are allowed by conservation of angular momentum, with $M = 0$, $M = 0, \pm2$ and $M = 0,\pm2$ and $\pm 4$ respectively. Additionally $\beta_{L-M} = (-1)^M\beta^{*}_{LM}$. This provides a total of only 6 independent equations. Even for one-photon excitation of only a single excited state, the selection rules for, and relations between the MADMs result in 4 independent MADMs two of which are complex, leaving 6 unknowns in the absence of any additional symmetry. Adding an additional excited state adds additional non-zero MADMs, but does not increase the number of unique $\beta_{LM}$ rendering the problem ill-posed.

Nonetheless, in certain cases all of these difficulties can potentially be overcome by symmetry, and limiting the number of excited states. One such case is the previously mentioned $B^{ 1}E''$ state of of ammonia. This is nominally a doubly degenerate electronic state, in which degeneracy is broken by Jahn-Teller distortion and the Electronic Coriolis coupling~\cite{allen1991}. In a recent experiment with linearly polarized ionizing light, it was shown that the various MADMs are separately projected into the anisotropy parameters $\beta_{00}$, $\beta_{20}$ and $\beta_{40}$, the coherence between the near-degenerate states being detected in $\beta_{40}$~\cite{makhija2020}. The Koopman's Correlations fail in this case and both electronic states ionize to the ground ionic state ejecting energetically degenerate photoelectrons. Further, symmetry of the electronic states imposes additional selection rules resulting in the excitation of only three unique MADMs, all of which are real~\cite{makhija2020}. We note that since the excitation pulse used in the experiment was significantly longer than the timescale of the ground state-excited state coherence, this coherence is washed out by the pulse. A complete experiment has also been performed through this state by REMPI~\cite{hockett2009}, therefore the $C^{LM}_{KQS}(n,n';\epsilon)$ can be determined. This then provides three independent equations from the combination of the time and frequency resolved experiments, resulting in sufficient information to determine the three unknown MADMs and thus the lab frame density matrix as a function of time. We are currently working towards this goal.

\section{Summary and Outlook}
We have developed a density matrix formalism for quantum molecular dynamics in terms of Molecular Angular Distribution Moments (MADMs) which elucidates the transformation from the molecular frame to the lab frame. This connection is clearly provided by Eq.~\ref{eq:PintermB}, the vibronic density matrix for a molecule at a particular orientation. Importantly, we find that the lab frame orientation of certain coherences - such as that between the ground and resonantly excited state - removes them from the experimentally accessible reduced vibronic density matrix. Nonetheless, these coherences may be accessible by measurement indicating that the orientation-averaged reduced vibronic density matrix is not an accurate representation of experimentally accessible information. The MADMs can therefore serve as an improved comparison with experimental data. Further, since the lab frame vibronic density matrix Eq.~\ref{eq:ADMOrigin} and Eq.~\ref{eq:EADMOrigin} is orientation dependent, in any given experiment initiated from an isotropic ensemble a distribution of vibronic dynamics will occur (cf. Fig.~\ref{fig:electrondensity}). This fact can be exploited to extract the MADMs from an experiment, and therefore the density matrix elements for any orientation angle, constituting a molecular frame tomography. This serves as the raw data to produce time resolved probability densities, such as those in Fig.~\ref{fig:electrondensity}, for a molecule at any orientation angle providing an experimental route to imaging molecular frame vibronic dynamics. Additionally, the ability to perform such a tomography is also the necessary first step toward the detailed experimental study of fundamental quantum mechanical phenomena in individual molecules, such as entanglement. Recent measurements have shown that exploiting attosecond laser pulses makes such phenomena accessible on the timescales of electronic motion~\cite{vrakking2021,koll2022}. We anticipate that quantum tomography will be a vital ingredient for further studies in this direction, as has been the case for such investigations with optical systems~\cite{dariano2003}.      

\newpage

\begin{appendices}

\section{Derivation of the Wigner D-Matrix Matrix Elements}
\label{appendix:dmatrix}
\renewcommand{\theequation}{A.\arabic{equation}}
\setcounter{equation}{0}
Using the resolution of the identity in the Euler Angle basis $1=\int{}\ket{\mathbf{\Omega}}\bra{\mathbf{\Omega}}d\mathbf{\Omega}$,
\begin{equation}
    \bra{J_nK_n M_n} D^{1*}_{0q} \ket{J_{in} K_{in}M_{in}}= \int{} \bra{J_n K_n M_n}\ket{\mathbf{\Omega}}D^{1*}_{0q} \bra{\mathbf{\Omega}}\ket{J_{in}K_{in}M_{in}} d\mathbf{\Omega}
\end{equation}
Then using Eq.~\ref{eq:JKMbasis} for the symmetric top wavefunction $\bra{\mathbf{\Omega}}\ket{J K M}$ gives
\begin{equation}
   \bra{J_nK_n M_n} D^{1*}_{0q} \ket{J_{in} K_{in}M_{in}}= \int{}\frac{\sqrt{(2J_n+1)(2J_{in}+1)}}{8\pi^2}D^{J_n}_{M_nK_n}D^{1*}_{0q}D^{J_{in}*}_{M_{in}K_{in}} d\mathbf{\Omega}
\end{equation}
Further simplification to this equation can be achieved by using the relationship $D^{j*}_{m'm}=(-1)^{m-m'}D^{j}_{-m'-m}$
\begin{equation}
\begin{split}
    \bra{J_nK_n M_n} D^{1*}_{0q} \ket{J_{in} K_{in}M_{in}}= \frac{\sqrt{(2J_n+1)(2J_{in}+1)}}{8\pi^2}(-1)^{q-0}(-1)^{K_{in}-M_{in}} \times\\ \int{}D^{J_n}_{M_nK_n}D^{1}_{0-q}D^{J_{in}}_{-M_{in}-K_{in}} d\mathbf{\Omega}
\end{split}
\end{equation}
The solution to this integral is well known and given in ~\cite{zare1988}. Thus, $\bra{J_nK_n M_n} D^{1*}_{0q} \ket{J_{in} K_{in}M_{in}}$ can be expressed in terms of a set of Wigner 3j symbols,
\begin{equation}
\begin{split}
    \bra{J_nK_n M_n} D^{1*}_{0q} \ket{J_{in} K_{in}M_{in}} = \sqrt{(2J_n+1)(2J_{in}+1)}(-1)^{q}(-1)^{K_{in}-M_{in}} \times \\ \begin{pmatrix} J_{in}& 1 &J_n \\ -M_{in} & 0 & M_n \end{pmatrix}\begin{pmatrix} J_{in}& 1 &J_n \\ -K_{in} & -q & K_n \end{pmatrix}
\end{split}
\end{equation}

\section{Alternative Derivation of the MADMs}
\label{appendix:madms}
\renewcommand{\theequation}{B.\arabic{equation}}
\setcounter{equation}{0}
We consider $\left\langle \mathbf{\Omega} n \right|\rho(t)\left| n \mathbf{\Omega} \right\rangle$, which is the probability of finding a molecule in vibronic state $n$ at the orientation $\mathbf{\Omega}={\phi, \theta, \chi}$. Using Eq.~\ref{eq:DDcombination} and \ref{eq:JKMbasis} in Eq.~\ref{eq:rho-op}, we find
\begin{equation}
\begin{split}
& \left\langle \mathbf{\Omega} n \right|\hat{\rho}(t)\left| n \mathbf{\Omega} \right\rangle = \sum_{KQS}\frac{(2K+1)}{8\pi^2}\sum_{J_n K_n M_n}\sum_{J'_{n} K'_{n} M'_{n}}\rho_{{\omega}_n {\omega}'_{n'}}(t) \\
&\times \sqrt{(2J_n+1)(2J_{n'}+1)}(-1)^{(M_n-K_n)}\left(\begin{array}{c c c} J_n & J'_n & K \\ -M_n & M'_n & Q\end{array}\right) \left(\begin{array}{c c c} J_n & J'_n & K \\ -K_n & K'_n & S \end{array}\right)D^{K*}_{QS}(\mathbf{\Omega}).
\end{split}
\end{equation}
The matrix elements of the AMCOs are easily found using Eq.~\ref{eq:AKQS}. Comparing the matrix element\\ $\left\langle J'_{n'} K'_{n'} M'_{n'} n' \right| A^K_{QS} \left| n J_n K_n M_n \right\rangle$ to the second line of the above equation for $\left\langle \mathbf{\Omega} n \right|\rho(t)\left| n \mathbf{\Omega} \right\rangle$ gives,
\begin{equation}
 \left\langle \mathbf{\Omega} n \right|\hat{\rho}(t)\left| n \mathbf{\Omega} \right\rangle = \sum_{KQS}D^{K*}_{QS}(\mathbf{\Omega})\sum_{J_n K_n M_n}\sum_{J'_{n} K'_{n} M'_{n}}\rho_{{\omega}_n {\omega}'_{n}}(t) \left\langle J'_n K'_n M'_n n \right| A^K_{QS} \left| n J_n K_n M_n \right\rangle .
\end{equation}
The double sum over rotational states above is the definition of the ADMs from Eq.~\ref{eq:EADMs}, 
\begin{equation}
 \left\langle \mathbf{\Omega} n \right|\hat{\rho}(t) \left| n \mathbf{\Omega} \right\rangle =  \sum_{KQS} A^K_{QS}(n,n,t) D^{K*}_{QS}(\mathbf{\Omega}),
\label{eq:axisdist}
\end{equation}
Which is identical to Eq.~\ref{eq:ADMOrigin}, the diagonal elements of the lab frame density matrix. An expression identical to Eq.~\ref{eq:EADMOrigin} for $\left\langle \mathbf{\Omega} n' \right| \hat{\rho}(t) \left| n \mathbf{\Omega} \right\rangle$ can be derived following the same procedure giving the EADMs and off-diagonal elements of the lab frame density matrix.

\section{Calculation of one-electron difference densities}
\label{appendix:diffdens}
\renewcommand{\theequation}{C.\arabic{equation}}
\setcounter{equation}{0}
The orientation- and time-dependent one-electron difference density $\Delta p(\boldsymbol{r}_{1}, \mathbf{\Omega}, t)$ (Eqaution \ref{eq:diffdens}) may be expressed in terms of the molecular frame electronic wave functions $\Phi_{\alpha}(\boldsymbol{r}_{1},\dots,\boldsymbol{r}_{N})$ and density matrix elements $\rho_{\alpha \alpha'}(\mathbf{\Omega}, t)$ as

\begin{equation}
\Delta p(\boldsymbol{r}_{1}, \mathbf{\Omega}, t) = \sum_{\alpha \alpha'} \rho_{\alpha \alpha'}(\mathbf{\Omega}, t) p_{\alpha \alpha'}(\boldsymbol{r}_{1}) - p_{00}(\boldsymbol{r}_{1}),
\end{equation}

\noindent
where $p_{\alpha \alpha'}(\boldsymbol{r}_{1})$ denotes the one-electron reduced (transition) densities,

\begin{equation}
p_{\alpha \alpha'}(\boldsymbol{r}_{1}) = N \int d\boldsymbol{r}_{2} \cdots d\boldsymbol{r}_{N} \Phi_{\alpha}^{*}(\boldsymbol{r}_{1},\dots,\boldsymbol{r}_{N}) \Phi_{\alpha'}(\boldsymbol{r}_{1},\dots,\boldsymbol{r}_{N}).
\end{equation}

In practice, the one-electron reduced (transition) densities are expanded in terms of a single-particle basis $\{ \varphi_{p}(\boldsymbol{r}_{1}) \}$,

\begin{equation}
p_{\alpha \alpha'}(\boldsymbol{r}_{1}) = \sum_{pq} d_{pq}^{(\alpha, \alpha')} \varphi_{p}^{*}(\boldsymbol{r}_{1}) \varphi_{q}(\boldsymbol{r}_{1}),
\end{equation}

\noindent
where the one-electron reduced (transition) density matrices $d_{pq}^{(\alpha, \alpha')}$ are defined as

\begin{equation}
d_{pq}^{(\alpha, \alpha')} = \left\langle \Phi_{\alpha} \middle| \hat{a}_{p}^{\dagger} \hat{a}_{q} \middle| \Phi_{\alpha'} \right\rangle,
\end{equation}

\noindent
with $\hat{a}_{p}^{\dagger}$ ($\hat{a}_{p}$) denoting the elementary fermionic creation (annihilation) operators corresponding to the single-particle basis $\{ \varphi_{p}(\boldsymbol{r}_{1}) \}$. The $\boldsymbol{d}^{(\alpha, \alpha')}$ were computed at the p-DFT/MRCI level of theory using the def2-TZVP basis.

\end{appendices}
\newpage
\selectlanguage{english}
\FloatBarrier
\bibliographystyle{unsrt}  

\bibliography{refs}

\end{document}